# Defect segregation and its effect on the photoelectrochemical properties of Ti-doped hematite photoanodes for solar water splitting

Barbara Scherrer*,+,[1], Tong Li*+,[2,3], Anton Tsyganok[1], Max Döbeli[4], Bhavana Gupta[5], Kirtiman Deo Malviya[4], Olga Kasian[3], Nitzan Maman[5], Baptiste Gault[3], Daniel A. Grave[1], Alexander Mehlman[1], Iris Visoly-Fisher[5], Dierk Raabe[3] and Avner Rothschild[1]

+shared first authorship

[1]Department of Materials Science and Engineering, Technion – Israel Institute of Technology, Haifa, Israel

[2] Institute for Materials & Zentrum für Grenzflächendominierte Höchstleistungswerkstoffe (ZGH), Ruhr-Universität Bochum, Bochum, Germany

[3] Max-Planck-Institut für Eisenforschung, Düsseldorf, Germany

[4]Ion Beam Physics, ETH Zurich, Zurich, Switzerland

[5]Department of Solar Energy and Environmental Physics, Swiss Institute for Dryland Environmental and Energy Research, Blaustein Institutes for Desert Research, Ben-Gurion University of the Negev, Midreshet Ben-Gurion, Israel

*corresponding author: barbara.scherrer@alumni.ethz.ch and tong.li@rub.de

## Abstract

Optimising the photoelectrochemical performance of hematite photoanodes for solar water splitting requires better understanding of the relationships between dopant distribution, structural defects and photoelectrochemical properties. Here, we use complementary characterisation techniques including electron microscopy, conductive atomic force microscopy (CAFM), Rutherford backscattering spectroscopy (RBS), atom probe tomography (APT) and intensity modulated photocurrent spectroscopy (IMPS) to study this correlation in Ti-doped (1 cat.%) hematite films deposited by pulsed laser deposition (PLD) on F:SnO$_2$ (FTO) coated glass substrates. The deposition was carried out at 300 °C, followed by annealing at 500 °C for 2 h. Upon annealing, Ti was observed by APT to segregate to the hematite/FTO interface and into some hematite grains. Since no other pronounced changes in microstructure and chemical composition were observed by electron microscopy and RBS after annealing, the non-uniform Ti redistribution seems to be the reason for a reduced interfacial recombination in the annealed films, as observed by IMPS. This results in a lower onset potential, higher photocurrent and larger fill factor with respect to the as-deposited state. This work provides atomic-scale insights into the microscopic inhomogeneity in Ti-doped hematite thin films and the role of defect segregation in their electrical and photoelectrochemical properties.

**Introduction**

With increasing energy demands, developing affordable and sustainable solar energy conversion and storage technologies becomes crucial. Photoelectrochemical (PEC) water splitting is an elegant approach to convert solar energy to clean hydrogen fuel, wherein sunlight is absorbed by semiconductor photoelectrodes to produce electron-hole pairs which reduce and oxidise water to hydrogen and oxygen, respectively [1]. Hematite (α-Fe$_2$O$_3$) has emerged as a leading photoanode candidate for PEC water splitting due to its favourable optical bandgap (~2.1 eV) [2] that corresponds to absorption of wavelengths shorter than ~600 nm, stability in neutral and basic aqueous solutions, and abundance [3]. With a theoretical potential to convert up to 16% of the incident solar energy to hydrogen, hematite photoanodes have attracted considerable attention and have been extensively examined [3-5]. Despite these efforts, the solar energy conversion efficiency that has been achieved with hematite photoanodes remains rather low [3], which was attributed to futile light absorption (*d-d* transitions) and fast charge carrier (electron-hole) recombination within the bulk and at the surface [6].

A common approach to improve the PEC performance of hematite photoanodes thus lies in modifying their properties via doping [3, 5, 7] . Extensive investigation using dopants such as $Mg^{2+}$ and $Cu^{2+}$ serving as acceptors, or $Ti^{4+}$, $Sn^{4+}$, $Zr^{4+}$, $Si^{4+}$, $Nb^{5+}$, $Pt^{4+}$ serving as donors have been reported [8]. In particular, Ti-doped hematite photoanodes demonstrated promising PEC properties toward water photo-oxidation, well above their undoped counterparts [8-14]. $Ti^{4+}$ replaces $Fe^{3+}$ substitutionally, introducing either electrons or iron vacancies depending on process conditions [15-18]. Ti-doping was found to reduce the unit cell volume of hematite and reduce the Fe-Fe bond length, which influences the hopping probability of localised charge carriers (polarons) [19]. Therefore, the carrier concentration and mobility do not scale linearly with Ti-doping concentration [20]. Ti doping levels of less than 1 cat.% were reported to exhibit improved PEC performance compared to higher doping levels (7 cat.%) [18]. Recently, a time resolved microwave conductivity study of charge carrier dynamics in hematite thin films has suggested that Ti-doping results in an increase in charge carrier mobility [21]. Although these studies have confirmed that Ti-doped hematite photoanode can improve the water photo oxidation performance [8-14], the exact Ti distribution at the atomic scale and the interplay between the dopant distribution and PEC properties remain unclear. Glasscock *et al.* [9] attributed the improved performance of Ti-doped thin film hematite photoanodes compared to their Si-doped counterparts to enhanced surface charge transfer kinetics, possibly due to the passivation of grain boundaries by Ti. Zandi *et al.* [11] proposed that Ti-doping possibly 'awakes' the 'dead layer' at the hematite/FTO interface which results in an activated surface for water oxidation and an improved hole collection length. However, no conclusive experimental data was reported to unveil the location of Ti in the photoanode.

Therefore, the mechanism by which Ti doping improves hematite photoanodes remains unresolved.

Besides doping, thermal treatment, e.g. annealing, was found to enhance the water photo-oxidation performance [22, 23]. Annealing in air at 800 °C was found to reduce the onset potential for water photo-oxidation, possibly due to the passivation of shallow surface states [22]. Li *et al.* [24] observed a significant improvement when $TiO_2$ coated hematite nanorods were annealed above 650 °C, with concomitant increase in both the short range order and surface defects observed by Raman spectroscopy. Sivula *et al.* [23] reported that the photoactivity of hematite photoanodes was drastically enhanced upon sintering at 800 °C, which was attributed to Sn inter-diffusion from the FTO substrate into hematite. The different observations made in these studies show that the effect of annealing on the microstructure and dopant distribution requires further investigation.

Here, we examine the dopant distribution and its influence on the electrical conductivity and photoelectrochemical properties of Ti-doped hematite thin film photoanodes prepared by pulsed laser deposition (PLD). PLD is known for its highly reproducible films with controlled microstructure and morphology that makes them well-suited for systematic investigations [25-27]. In order to investigate the correlation between PEC properties, microstructure and defect distribution, both within the bulk, at the hematite/FTO interface and the surface, complementary characterisation techniques were employed including scanning electron microscopy (SEM), transmission electron microscopy (TEM), conductive atomic force microscopy (CAFM), Rutherford backscattering spectrometry (RBS), and atom probe tomography (APT), in conjunction with photoelectrochemical measurments. This correlative analysis provides new insights into the influence of annealing and dopant distribution at the atomic scale on photoelectrochemical properties.

**Experimental**

Deposition

Deposition of the hematite films was conducted by PLD at a set point temperature of 300 °C from self-made targets, at a pressure of 3.33 Pa in an oxygen atmosphere [8]. The distance between the target and the substrate was 7.5 cm [8]. The annealing treatment was carried out at 500° C for 2 h in air at a heating and cooling rate of 10° C/min. Glassy carbon (HTW Hochtemperatur-Werkstoffe GmbH), silicon wafer pieces (Universal wafer) and fluorinated tin oxide (FTO) coated soda-lime glass (TEC15, Pilkington) were used as substrates. The FTO substrates were cleaned as described in reference [25]. The deposition of the hematite films was performed in a PLD workstation (SURFACE systems+technology GmbH & Co. KG) with a KrF Excimer Laser at 3 Hz and 1 J/cm$^2$.

Characterisation

Photoelectrochemical measurements were carried out in "Cappuccino cells" electrochemical test cells equipped with a potentiostat (CompactStat, Ivium Technologies) and a solar simulator (Sun 3000 class AAA solar simulator, ABET Technologies, AM 1.5G) as a light source [28]. All measurements were carried out in 1 M NaOH aqueous solution (pH 13.6). The current was measured as a function of the electrode potential using a 3-electrode setup with an Hg/HgO reference electrode and a platinum counter electrode at a scanning rate of 10 mV/s. The applied potential was converted to the RHE scale using the Nernst equation.

IMPS analysis was performed using Zahner CIMPS system equipped with white light LED source (4300 K). Photo-electrochemical impedance spectroscopy (PEIS) and intensity modulated photovoltage spectroscopy (IMVS) spectra were directly measured to calculate better quality IMPS, as described elsewhere [29]. Light intensities of 50 and 100 mW/cm$^2$ were used during the data acquisition process, by applying 0.8 V or 1.6 V DC voltage on the light source power supply. AC perturbation of 20 mV was superimposed on the LED power supply to perform IMVS measurements. For PEIS measurements AC modulation of 10mV was superimposed on the DC potential bias. Frequency range of 300 mHz-10 kHz was used for all of the described measurements.

XRD analysis was performed using a Rigaku SmartLab X-ray diffractometer. The acquisition conditions were parallel beam configuration with Cu Kα radiation source ($\lambda$ = 0.15406 nm) in the 2θ range of 20−75° at a scan rate of $0.01°s^{-1}$.

Topography and current mapping were conducted by atomic force microscope (AFM, Keysight 5500) in contact mode, using a Pt-coated probe (ANSCM-PC). Ag paste (Ted Pella, inc.) was applied to the Si as a back contact. The bias for current mapping (2 V or -2 V) was applied to the Si substrate while the probe was held at virtual ground. The measurements were performed at room temperature. In order to estimate the conductive area % in current mapping obtained through CAFM, we used image analysis in the software WSxM 5.0 [30]. The conductive area % is estimated by using a fixed threshold of > 5 pA.

The SEM images were recorded by an inlens detector in a Zeiss Ultra-Plus FEG-SEM at an accelrating voltage of 4 kV. Thin lamelas for TEM cross-section observation were prepared using a FEI Helios 600 Nanolab focused ion beam (FIB) with a micromanipulator (Omniprobe 200). The TEM analysis was performed using a monochromated and aberration (image)-corrected TEM (FEI Titan 80–300 TEM). STEM analyses were carried out at 300 kV with 0.240 nA current and a convergence semiangle of 10 mrad.

The composition as a function of depth of hematite-$SnO_2$ specimens deposited on glassy carbon and silicon were analysed by RBS using a 2 MeV and 5 MeV 4He beam and a silicon

PIN diode detector at 168°. The experimental data were fitted using RUMP, assuming that there is stoichiometric oxygen [31].

Needle-shaped APT specimens were prepared by means of a site-specific lift-out procedure using a FEI Helios G4 CX focused ion beam (FIB)/scanning electron microscope. A 200 nm thick protective Cr-layer was coated on top of hematite by electron-beam deposition in order to protect the surface. The APT experiments were conducted on a CAMECA LEAP 5000 XR instrument equipped with an ultraviolet laser with a spot size of 2 µm and a wavelength of 355 nm. The detection efficiency of this state-of-the-art microscope is ~52%. Data was acquired in laser pulsing mode at a specimen temperature of 60 K, with a target evaporation rate of 5 ions per 1000 pulses, a pulsing rate of 200 kHz, laser pulse energy of 60 pJ. The APT data were reconstructed and analysed using the commercial IVAS 3.8.2 ™ software.

## Results and discussion

Hematite thin films (~40 nm) were deposited on FTO-coated glass substrates by PLD at 300° C, and subsequently annealed at 500 °C for 2 h in air. Depositions at temperatures higher than 300 °C resulted in cracking of the hematite thin film (see Figure S1). The XRD diffractograms of the photoanodes before and after annealing in Figure 1a display Bragg reflections of FTO and hematite, indicating that heamtite thin films were deposited and both as-deposited and annealed hematite thin films are crytalline. The microstructure was examined by SEM and TEM before and after annealing (Figure 1b-i, S2). No cracks are observed in either the as-deposited state or after annealing, as observed in Figures 1b-c. Note that these SEM images were taken from a hematite thin film deposited on a flat silicon wafer instead of an FTO-coated glass substrate as in the photoanodes because it is difficult to observe the microstructure of hematite thin films on FTO-coated glass substrates due to the roughness of the FTO layer (see Figure 1d-e). The deposition conditions on silicon wafer and FTO-coated glass substrate were kept the same and the observed microstructure and crystal structure (hematite) of the thin films were found to be independent of the substrate materials under the chosen deposition conditions (Figure 1). Cross-section TEM images of the hematite films are presented in Figures 1f-i and S2. The high-angle annular dark-field scanning transmission electron microscopy (HAADF-STEM) images in Figure 1f-g and S2 reveal that the 40 nm-thick hematite film (grey contrast) coats conformably the rough surface of the FTO layer (bright contrast). High-resolution TEM (HRTEM) images at higher magnification, as shown in Figure 1h-i, futher confirm that the hematite films before and after annealing are crystalline, which is consitent with the XRD data (Figure 1a). The XRD, SEM and TEM results reveal that both as-deposited and annealed hematite thin films are crystalline and crack-free. No noticeable difference in the microstructure can be observed in the as-deposited films vs. the annealed films.

The dashed curves show the photocurrent, that is the light current subtracted by the dark current. The photoelectrochemical properties toward water photo-oxidation were examined by linear sweep voltammetry (LSV) in dark and under solar-simulated illumination, carried out in 1 M NaOH solution in deionised water. The LSV voltammograms of photoanodes in the as-deposited state and after annealing are shown in Figure 2a. After annealing, the photoelectrical performance was improved considerably, displaying lower onset potential, higher photocurrent and larger fill factor with respect to the as-deposited photoanode. The onset potential of the annealed photoanode is slighly below 1.1 V vs RHE, more than 100 mV lower than that of the as-deposited photoanode. This observation suggests that annealing improves the competition between water photo-oxidation and electron – hole recombination at the front surface and/or the back interface [32]. In addition to the improvement in the charge carrier dynamics at the surface/interface that yields a lower onset potential, the annealing treatment also seems to improve bulk properties, giving rise to enhanced photocurrent at high potentials. In order to examine these hypotheses, intensity modulated photocurrent spectroscopy (IMPS) measurements were carried out in another set of photoanodes prepared by the same recipe. The IMPS spectra were analysed to extract the hole current ($J_h$, positive), recombination current ($J_r$, negative) and total photocurrent ($J = J_h + J_r$), as reported elsewhere [29, 32, 33]. The results, presented in Figure 2b, show that after annealing the magnitude of the recombination current ($J_r$) reduces significantly with respect to the as-deposited state, whereas the hole current ($J_h$) is slightly reduced. The significant reduction in interfacial recombination at the surface or/and interface with the substrate overcomes the small reduction in the hole current, yielding a higher photocurrent as well as a cathodic shift in the onset potential.

In order to examine the effect of annealing on the electrical conductivity in the dark, current and topography mapping of the photoanode were carried out by CAFM in the as-deposited state and after annealing, Figures 3a-c and S3-4. A bias of – 2 V or + 2 V was applied, where the current resulting from the positive bias indicates the ability to conduct electrons from the probe through the hematite to the back interface, or holes in the opposite direction. The measured current at + 2 V was up to 5-10 pA for the as-deposited photoanode, and it was strongly dependent on the location of the measurement. The grains were considerably more conductive than the grain boundaries that displayed near-zero currents. The current-voltage (I-V) curves were rectified, displaying Schottky-like nonlinear characteristics for both the grains and grain boundaries (Figure 3c). After annealing the conductivity of some grains was enhanced (classified as A-type grains), reaching currents of ~15 to ~50 pA at + 2 V (Figure 3b and S3f). About 20% of the surface area of the annealed photoanode displays enhanced conductivity relative to the as-deposited state. The more conductive grains (A-type grains in Figure 3c) display non-rectifying linear (Ohmic) I-V behaviour. 80 % of the area showed similar conductivity as in the as-deposited state, and these grains are termed B-type grains (The individual measurements are shown in Figure S4).

In order to futher examine the positive effect of the annealing on the electrical and photoelectrochemical properties, the chemical composition of the film was investigated in detail. The composition depth profile was investigated by RBS (Figure 4 and Figure S5). The RBS analysis was performed on hematite films that were deposited on glassy carbon and Si substrates since it is crucial to have a flat stratified structure for an accurate analysis. On both substrates, a $SnO_2$ interlayer was added between the substrate and the hematite film in order to simulate the FTO layer in the photoanode. The composition of the as-deposited hematite films was found to be $Fe_{1.93}Ti_{0.02}O_{3.0}$, obtained by fitting the RBS spectra of hematite films using RUMP [31] (see Figure 4a-b and Figure S5). The Ti to Fe ratio is close to the expected composition of the 1 cat.% Ti-doped $Fe_2O_3$ target, whereas the Fe content was lower than the stoichiometric composition, indicating the presence of Fe vacancies in the film [34, 35]. After annealing the Fe content dropped from 1.93 down to 1.90, indicating an increase in Fe vacancy during the annealing process. The shoulder to the right of the Sn peak (Figure 4d) may suggest a small amount of Sn within the hematite film. However, the quantitative analysis of this shoulder using RUMP results in $0.0002 \pm 0.0002$ cat.% Sn in the as-deposited state (Figure 4b) and $0.0004 \pm 0.0002$ cat.% Sn after annealing (Figure 4d). Since these values are very close to the detection limit of RBS (0.0007 % [36]), the presence of Sn within the hematite films cannot be ascertained. Furthermore, the small shoulder may arise from scratches, micro cracks or pinholes in the hematite film (such a pin hole is also observed in Figure S6a). This leads to the conclusion that, the RBS results show no conclusive evidence for Sn inter-diffusion into the hematite films, both in the as-deposited state and after annealing.

Our CAFM results (Figure 3) unveil that the conductivity (in the dark) of the annealed film is inhomogeneous, with enhanced conductivity and Ohmic behaviour observed in A-type grains. The high-resolution TEM results (Figure 1h-i) exclude the possibility that the inhomogeneous conductivity in our film results from heterogeneity in the crystal structure. Likewise, the XRD results (Figure 1a) do not show evidence for different orientations before and after annealing that might have affected the conductivity and photoelectrochemical properties, as reported for hematite layers deposited by atmospheric pressure chemical vapour deposition (APCVD) [37]. These results suggest that another factor, such as the local chemical composition, leads to locally enhanced conductivity. The RBS results can only provide the average chemical composition in the film, but not lateral distribution of elements within the film. Therefore, we carried out atom probe tomography (APT) analyses to examine the chemical composition at near-atomic scale in three dimensions [38-41]. In Figures 5a-b and 6a-b, the orange and blue regions indicate the hematite and FTO layers of the samples before and after annealing, respectively. The top surface of both 3D-APT reconstructions represents the surface of the hematite film as a protective deposition Cr layer (not shown) is present above the hematite layer. The hematite/Cr interface was identified by a sudden change in the voltage required for evaporating the Cr layer to that required by the hematite for the APT reconstruction. The cross-sectional elemental maps of Fe, Sn, O, Ti and C, cropped from the dashed rectangular box in Figure 5a and Figure 6a, show the elemental distribution in the hematite and FTO and the hematite/FTO interface (see Figure S7 and S8). Some region of the hematite film in the as-deposited state are enriched with oxygen (Figure 5b), which is confirmed and visualised by the O iso-concentration surface at 53.0 at.% (turquoise) in Figure 5c and the concentration profile in Figure 7a. In order to examine the correlation of these O-rich regions with Ti distribution, the Ti iso-concentration surface (red) is plotted in Figure 5d. It shows no correlation between the O-rich regions and Ti in the hematite of the as-deposited sample. After annealing, the O-rich grains remain in the hematite thin film (Figure 6b). Additionally, carbon was found to segregate to grain boundaries [42], as marked by grey dots in the cross-section view in Figure 6c and Figure S8. The distribution of carbon is therefore an indicator for grain boundaries. Based on this, the grain boundaries are marked by the green dashed lines in Figure 6d. Interestingly, the O-rich grain, highlighted by the O iso-concentration surface at 53.0 at.% in Figure 6c, is also enriched with titanium, as shown by the Ti iso-concentration surface in Figure 6d. One of the Ti-rich grains is found at the surface region of the hematite layer, indicating that the surface of the thin film is locally enriched with titanium. In addition to Ti-rich grain on the surface, titanium was observed to segregate at the hematite/FTO interface (Figure 6d) whereas such segregation was not seen in the as-deposited sample (Figure 5d).

The oxygen and titanium content in the O-rich grains before and after annealing was measured by the 1D concentration profiles in Figure 7a-b. The average cation ratio calculated from two O-rich grains with a size of ~30-40 nm and four O-lean grains all with a smaller grain size 20-30 nm in the annealed film, as summarised in Table S1, shows that the O-rich grains have ~1.6 cat.% Ti, whereas 0.5 cat.% Ti is present in the O-lean grains. In comparison, the O-rich grains in the as-deposited hematite thin film have ~0.4 cat.% Ti, and ~1.0 cat.% for the O-lean grains. After annealing, the Ti segregation at the hematite/FTO interface is ~ 3 cat.%, thereby lowering the interfacial energy and forming a more thermodynamically stable interface. Note that the oxygen content in both hematite and FTO (shown in Figures 7a-b and S6b) is lower than the stoichiometric value due to desorption of neutral oxygen atoms that fail to be detected by APT [43, 44]. Despite this, the Ti/Fe cation ratio should not be influenced.

We also examined the redistribution of Sn in the hematite film after annealing, following previous suggestions that Sn diffuses into hematite at high temperatures [23]. Mass spectra of two regions (indicated by the red circles in Figure 6b) in close proximity to the hematite/FTO interface are shown in Figures 7c-d. The local mass spectrum in hematite (Figure 7c) shows that no Sn signal was observed above the noise level, with a detection limit in the range of ~100 ppm, indicating that Sn is absent in the hematite layer. Additionally, Fe did not diffuse into FTO, Figure 7d. Sivula *et al.* concluded that Sn diffuses into hematite films from the FTO substrate during sintering at 800 °C, increasing the photocatalytic activity of hematite [23]. However, the hematite films investigated here were annealed at a much lower temperature (500 °C) which seems to be too low for Sn inter-diffusion. Therefore, the contribution of Sn diffusion at the hematite/FTO interface to the increase in the photoelectrochemical performance can be excluded in our study.

PLD is often considered to be a thin film deposition technique that transfers the target stoichiometry into the film evenly and accurately [45]. Our RBS results show that the ratio of Fe:Ti was transferred quite accurately to the films. However, the cation to oxygen ratio is nonstoichiometric as it is more dependent on the oxygen partial pressure during deposition then on the oxygen content of the target. This was also reported in previous studies of other materials deposited by PLD such as $SrTiO_3$ [46]. In addition to this, our APT data (Figure 5b) reveals that the O-rich grains are present in the as-deposited state. Upon annealing, Ti within the hematite film was found by APT to segregate to the O-rich grains due to the lower electron negativity $Ti^{4+}$ compared to $Fe^3$ (Figure 6d). Such non-uniform Ti distribution results in sub-stoichiometric cation (Fe + Ti) to anion (O) ratio. The localised sub-stoichiometric composition in these microscopic features introduces Fe vacancies ($V_{Fe}^{///}$) and Ti enrichment within these sub-stoichiometric features, as shown in the APT results presented in Figures 6c and d. According to the defect chemistry model of hematite, Fe vacancies are negatively charged ($V_{Fe}^{///}$) whereas Ti dopants on substitutional Fe sites are positively charged ($Ti_{Fe}^{\bullet}$) [15, 16, 34]. The positively charged Ti dopants (donors) compensate, at least partially, the negatively charged Fe vacancies that are prevelant in these regions. This gives rise to a non-uniform dopant distribution that could result in grains with different electrical conductivities in the dark, as observed in the CAFM conductivity maps (Figure 3b).

Interestingly, the hole current ($J_h$) during illumination was slightly reduced after annealing (Figure 2b), suggesting that the more conductive grains observed by CAFM in the annealed film might have a lower hole current than their less conductive counterparts. This hypothesis could be rationalized by the fact that higher electron concentration gives rise to faster electron – hole recombination, but further investigation needs to be carried out to validate this hypothesis. The significant reduction in interfacial recombination in the annealed film (Figure 2b) overcomes the small reduction in the hole current, resulting in a higher photocurrent and lower onset potential with respect to the as-deposited state (Figures 2). Given that no pronounced change in the microstructure and chemical composition were observed by XRD, SEM, TEM and RBS after annealing, the improvement of photoelectrochemical properties is therefore attributed mainly to Ti segregation at the hematite/FTO interface (Figure 6d) and possibly Ti-rich grains on the surface of annealed hematite film (Figure 6c) that lead to reduced interfacial recombination in the annealed films. Ti-segregation to the interface was previously suggested to activate the underlayer formed at the hematite/FTO interface, although no direct evidence of Ti segregation was provided [11]. Here we provide direct evidence of that claim. These results reveal another level of complexity in the microstructure – photoelectrochemical properties relationship in hematite photoanodes, beyond the crystallographic orientation [47] and high-angle grain boundaries [37] that were reported previously.

**Conclusions**

A combination of photoelectrochemical and microstructural analyses including XRD, SEM, HRTEM, CAFM, RBS and APT was employed to study how annealing improves the photoelectrochemical performance of Ti-doped hematite films deposited by PLD. The APT results indicate that the films have microscopically inhomogeneous distribution of O, which induces Ti segregation during annealing. In addition to this, Ti segregates to the hematite/FTO interface upon annealing. Given that no other profound changes in microstructure and composition were observed before and after annealing, the Ti redistribution is thought to be the primary reason for reduced interfacial recombination in the annealed films, as observed by IMPS, thereby improving the photoelectrochemical properties of the annealed films. These observations of microscopic inhomogeneity in chemical composition shed new light on the microstructure – photoelectrochemical properties relations in hematite photoanodes. Better understanding of the relationship between the non-uniform distribution at defects within hematite photoanodes and their electrical and photoelectrochemical properties opens a new pathway toward the design of next generation photoanodes for water splitting. Further optimisation efforts to improve the photoelectrochemical performance of such photoanodes should take these effects into account.

**Acknowledgements**

The research leading to these results has received funding from the Ministry of Science and Technology of Israel (grant no. 3-14423). B. Scherrer acknowledges support by Marie-Sklodowska-Curie Individual Fellowships no. 656132. T. Li thanks the Alexander von Humboldt Foundation for a postdoctoral fellowship and acknowledges funding from the Deutsche Forschungsgemeinschaft (DFG, German Research Foundation) – Projektnummer 388390466 – TRR 247 (C4). B. Gupta is grateful to the Blaustein Center for Scientific Cooperation (BIDR, BGU) for a postdoctoral fellowship. Some of the experiments reported in this work were carried out using central facilities at the Technion's Hydrogen Technologies Research Laboratory (HTRL) supported by the Adelis Foundation and the Solar Fuels I-CORE program of the Planning and Budgeting Committee and the Israel Science Foundation (Grant No. 152/11). The authors acknowledge support by the electron microscopy center of Technion (MIKA).

**Contributions**

B.S. planed the research, fabricated the samples and analysed them electrochemically, optically and by SEM. T.L. analysed the samples with APT and did the FIB work. B.S., T.L. and A.R. wrote the manuscript. B.G., N.M. and I.V-F. performed the AFM analysis and K.M. the TEM analysis. A.T. performed IMPS measurements and D.A.G. performed some photoelectrochemical measurements. A.M. prepared some of the hematite photoanodes. A.R. provided conceptual advice. All authors discussed the results and their interpretation as well as critically revised the manuscript.

**Competing financial interests**

The authors declare no competing financial interests.

**Supporting information**

High temperature deposition

Background information on CAFM

Background information on RBS

Background information on APT

# Figures

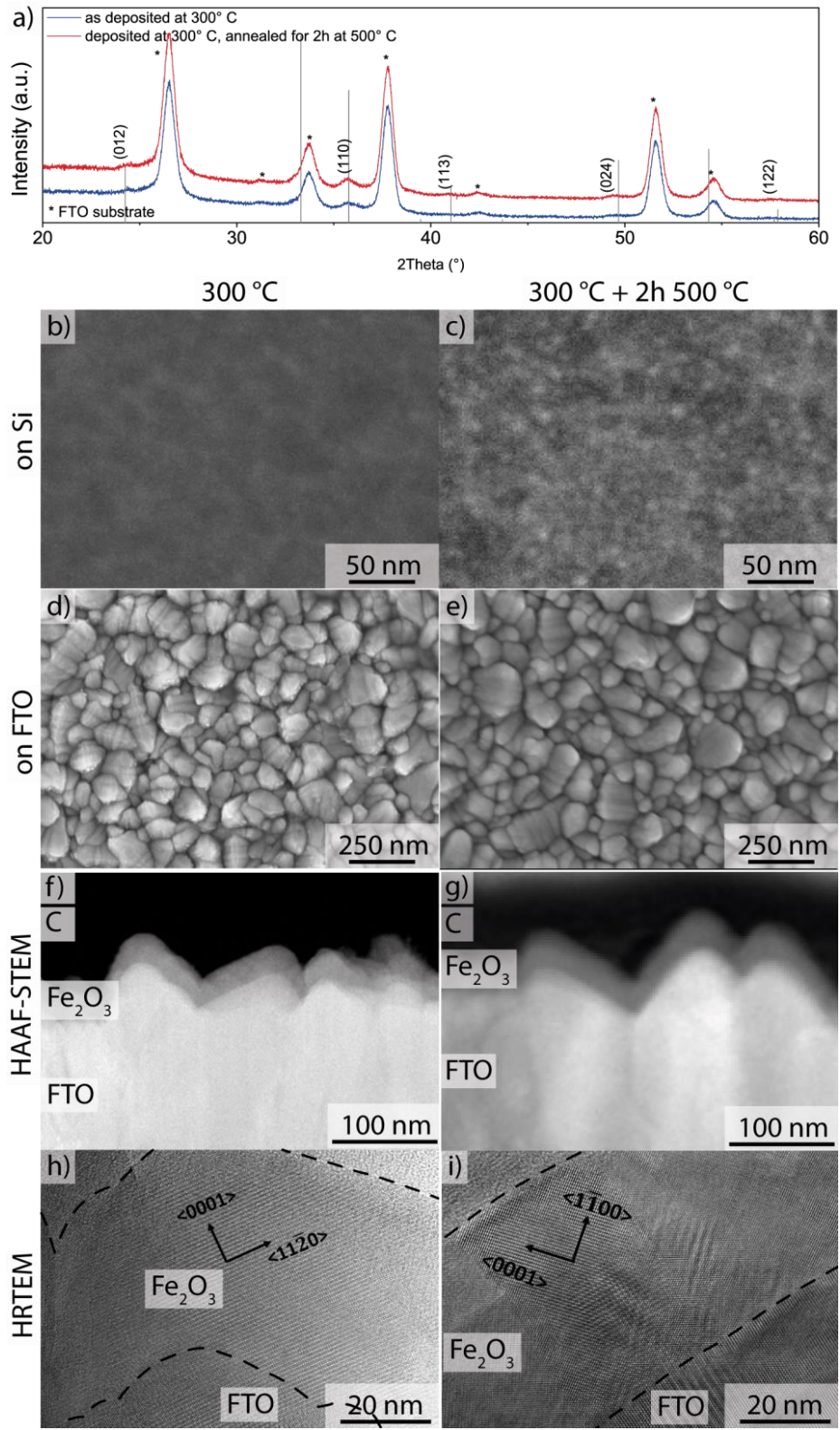

Figure 1. a) X-ray diffractograms of hematite films deposited on FTO-coated glass sbsutrates in the as-deposited state (blue curve) and after annealing (red curve). The Bragg reflection of hematite PXRD (from JCPDS 01-080-5406) are shown by the grey vertical lines. SEM images of b-c) as-deposited and annealed ~40 nm thick hematite film on Si wafer; d-e) as-deposited and annealed hematite film on FTO-glass; f-g) HAADF-STEM images of as-deposited and annealed hematite film, and h-i) HRTEM image of the as-deposited and annealed hematite film.

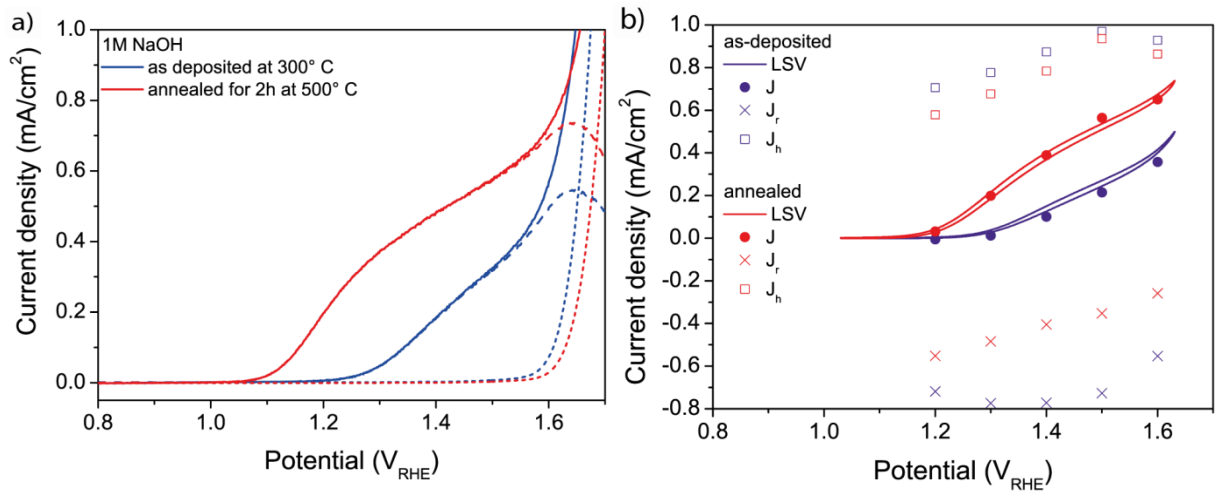

Figure 2. a) Linear sweep voltamograms measured in 1 M NaOH aqueous solution measured in the dark (dotted curves) and under solar-simulated illumination (solid curves). The dashed curves show the photocurrent, that is the light current subtracted by the dark current. b) Hole current ($J_h$, □), recombination current ($J_r$, ×) and total photocurrent ($J = J_h + J_r$, •) obtained from IMPS measurments of another set of photoanodes prepared and measurmed under the same conditions as in Figure 1a. The solid curves show the photocurrent measured by LSV. Blue and red curves correspond to the hematite photoanode in the as-deposited state and after annealing, respectively.

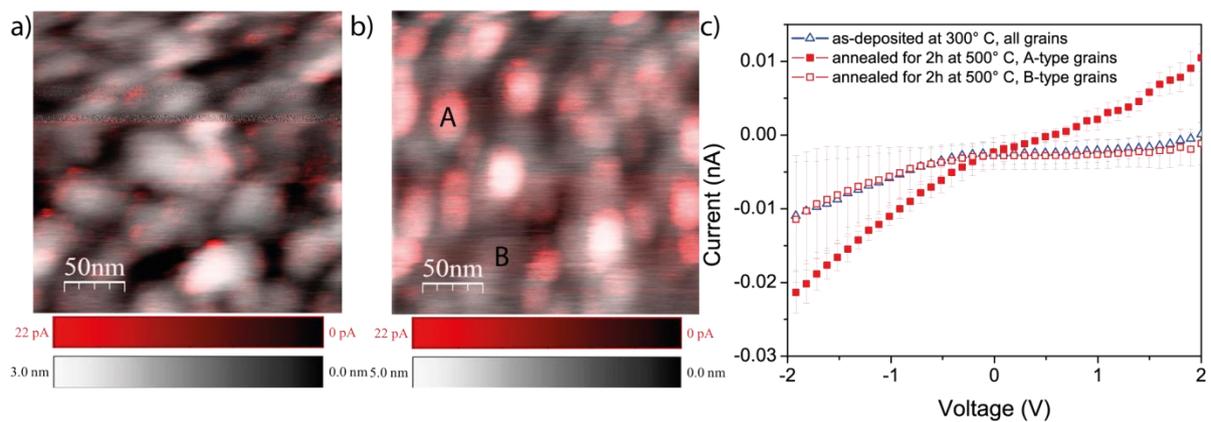

Figure 3. CAFM mapping (at 2 V, red scale) overlaid with topographical AFM mapping (grey scale) of a hematite film deposited on Si wafer in (a) the as-deposited state, and (b) after annealing for 2 h at 500° C; c) average current - voltage (I-V) curves measured locally on selected grains in the as-deposited hematite film (blue curve) and after annealing (red curves). Different grains in the as-deposited sample exhibit similar I-V curves (blue curve in c), whereas two distinct types of I-V curves marked by full and empty red squares were observed for A-type grains and B-type grains in the annealed sample. The individual measurements are shown in Figure S3.

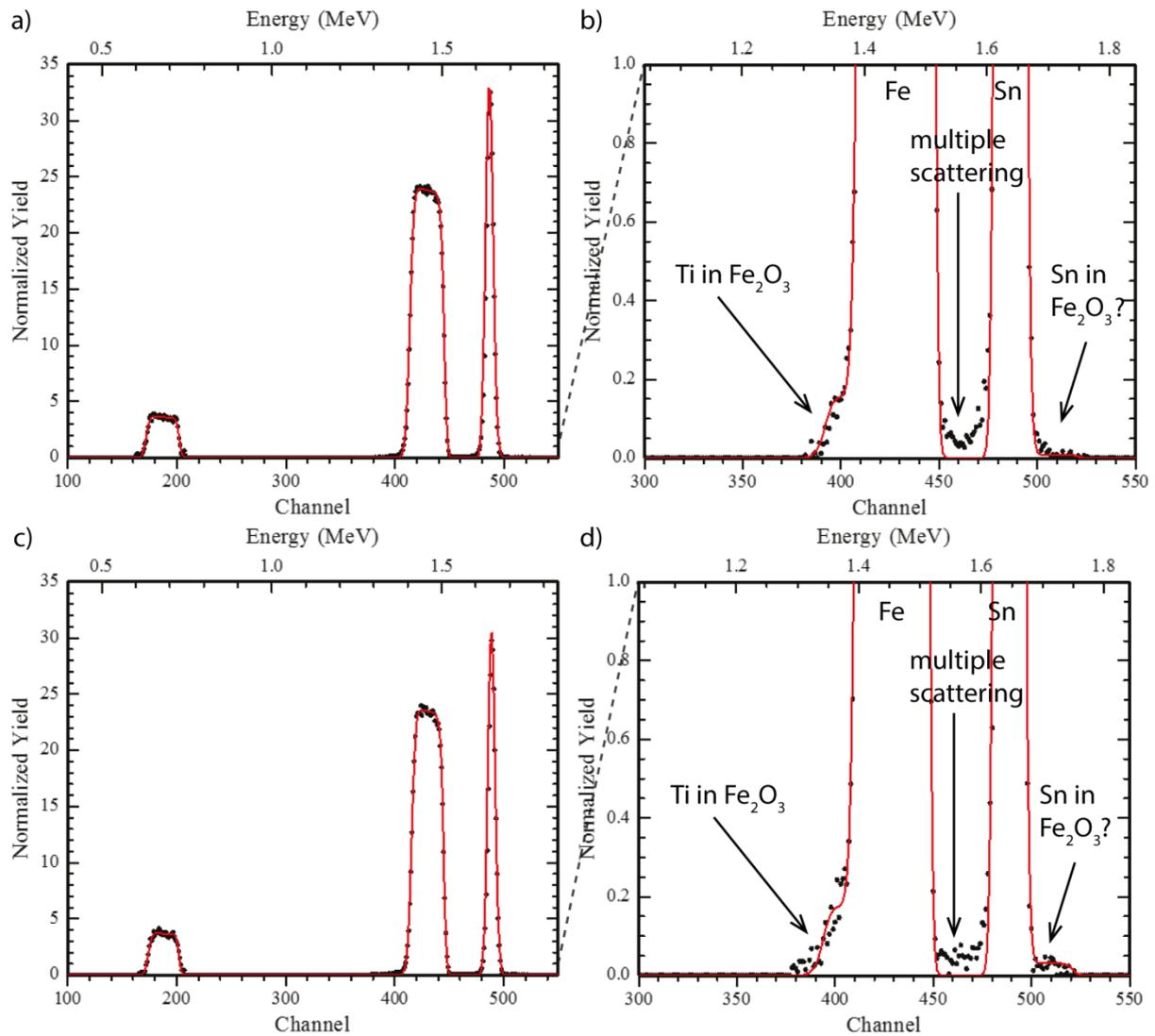

Figure 4. Overview and close-up Rutherford back scattering (RBS) spectra of hematite film with a tin oxide underlayer deposited on Si wafer: a) and b) as-deposited hematite at 300 °C, c) and d) after annealing at 500 °C for 2 h. Black dots are the measurement points and the red line is the fitted line.

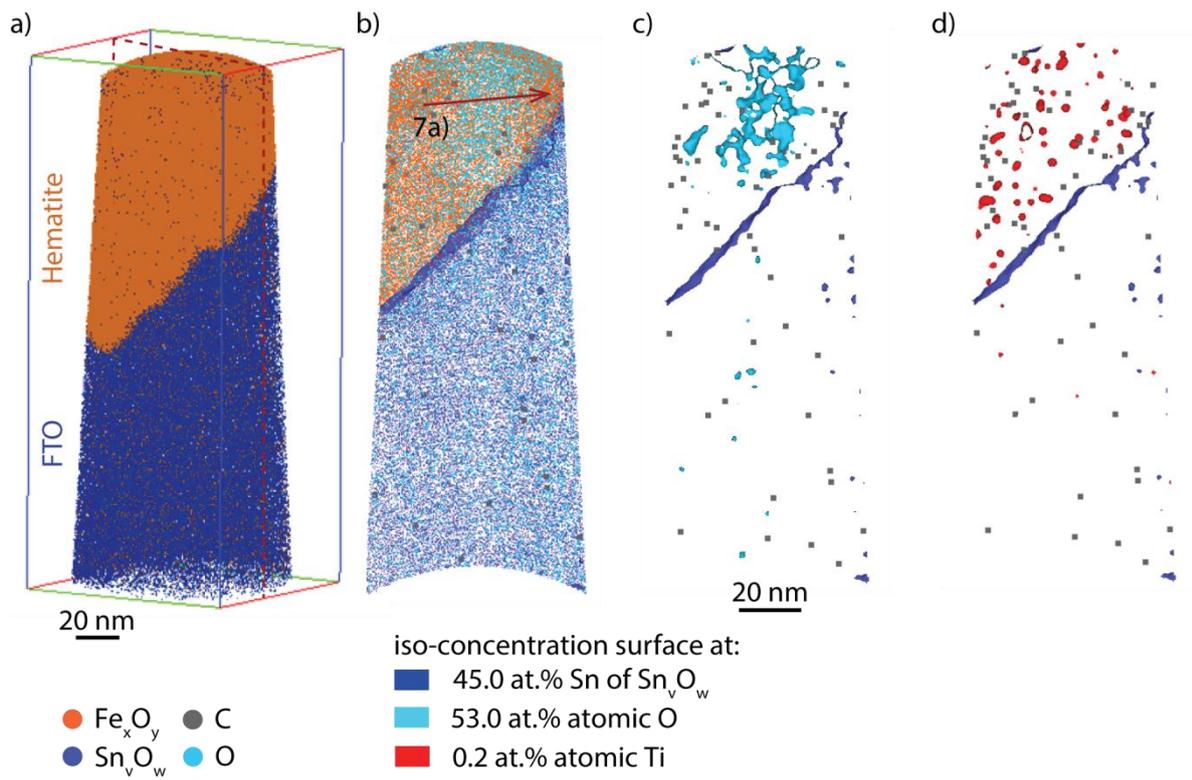

Figure 5: a) 3D-APT reconstruction of the hematite thin film deposited on FTO at 300 °C, b-d) cross-section of the APT reconstruction with iso-concentration at 45.0 at.% Sn of $Sn_vO_w$ (blue), 53.0 at.% O (turquoise) and 0.2 at.% Ti (red), respectively.

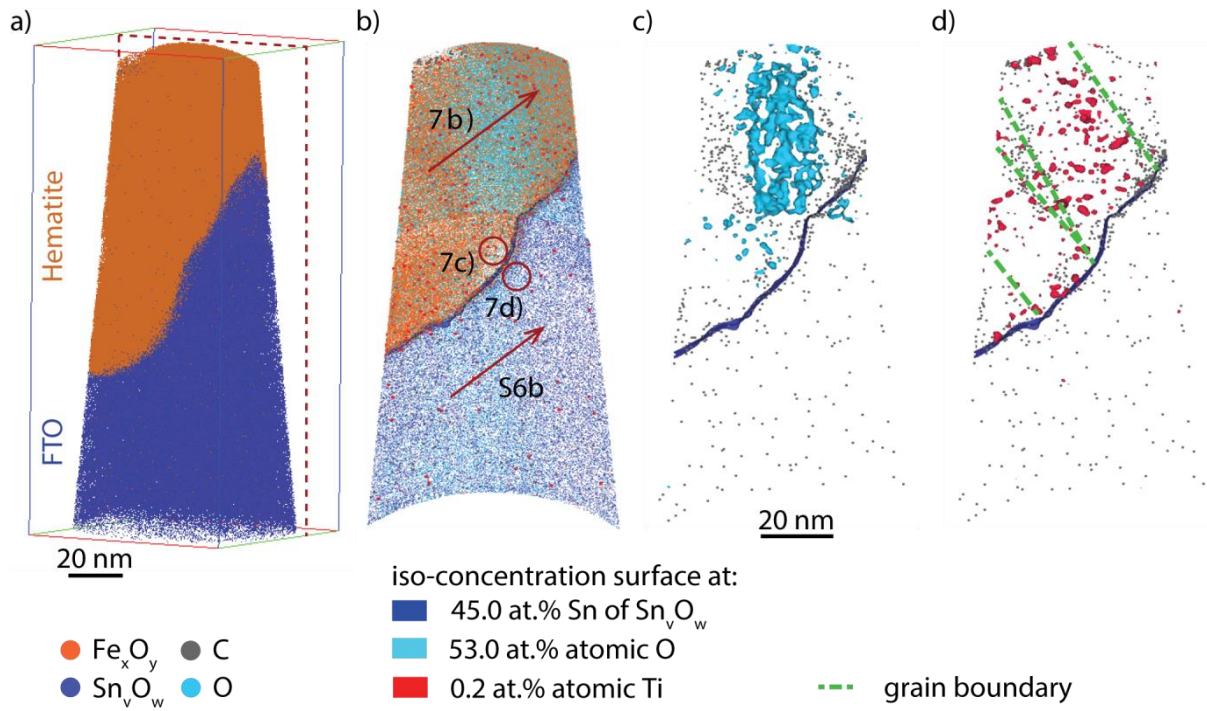

Figure 6: a) 3D-APT reconstruction of the hematite thin film deposited on FTO at 300 °C and annealed for 2 h at 500 °C, b-d) cross-section of the APT reconstruction with iso-concentration at 45.0 at.% Sn of $Sn_vO_w$ (blue), 53.0 at.% O (turquoise) and 0.2 at.% Ti (red), respectively.

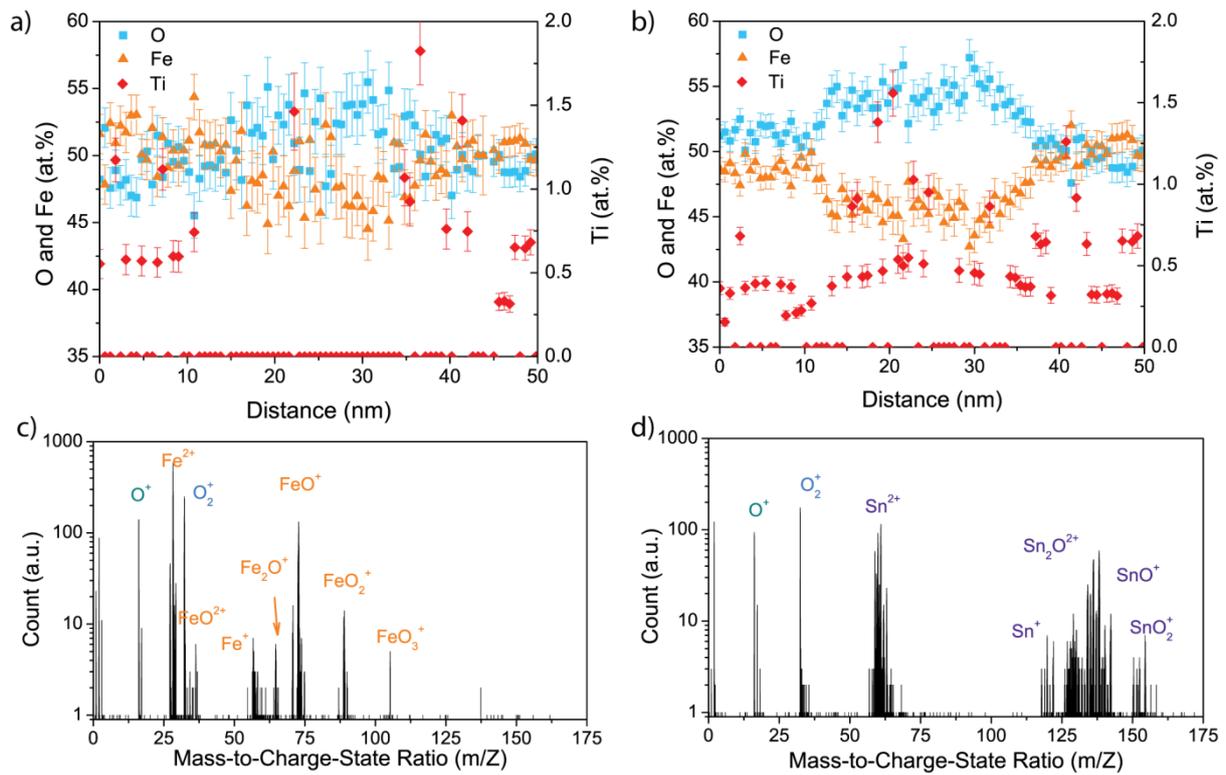

Figure 7: a-b) 1D concentration profiles taken from an analysis cylinders across O-rich regions in hematite indicated in Figure 5b and Figure 6b, c) and d) mass spectra from two regions in close proximity to the hematite/FTO interface, as indicated in Figure 6b.

# Table of Contents

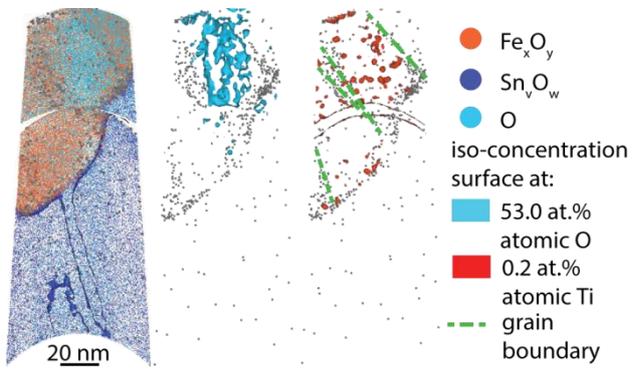